# The seeds and homogeneous nucleation of photoinduced nonthermal melting in semiconductors due to self-amplified local dynamic instability

Wen-Hao Liu[1,2], Jun-Wei Luo[1,2]*, Shu-Shen Li[1,2], and Lin-Wang Wang[3]*

[1]*State Key Laboratory of Superlattices and Microstructures, Institute of Semiconductors, Chinese Academy of Sciences, Beijing 100083, China*

[2]*Center of Materials Science and Optoelectronics Engineering, University of Chinese Academy of Sciences, Beijing 100049, China*

[3]*Materials Science Division, Lawrence Berkeley National Laboratory, Berkeley, California 94720, United States*

*Email: jwluo@semi.ac.cn; lwwang@lbl.gov

**Abstract:**

Laser-induced nonthermal melting in semiconductors has been studied over the last four decades, but the underlying mechanism is still under debate. Here, by utilizing an advanced real-time time-dependent density functional theory simulation, we reveal that the photoexcitation-induced ultrafast nonthermal melting in silicon occurs via homogeneous nucleation with random seeds originating from a self-amplified local dynamic instability at the photoexcited states rather than by simultaneously breaking of all bonds, as suggested by the inertial model, phonon instability, or Coulombic repulsion mechanisms. Due to this local dynamic instability, any initial small random thermal displacements of atoms can be amplified by a charge transfer of photoexcited carriers, which in turn creates a local self-trapping center for the excited carriers and yields the random nucleation seeds. Because a sufficient amount of photoexcited hot carriers must be cooled down to the band edges before participating in the self-amplification of local lattice distortions, the time needed for the cooling of this segment of hot carriers (rather than electron–lattice equilibration) is the response for the longer melting timescales at shorter laser wavelengths. This finding provides fresh insights into photoinduced ultrafast nonthermal melting.

**Introduction**

    Ultrashort laser pulses are currently used to manipulate the structure and function of materials at far from equilibrium states [1-9], with the corresponding ultrafast dynamics being one of the ultimate problems in modern science and technology. In terms of applications, the femtosecond and nanosecond pulsed laser was first utilized to deal with the annealing of the amorphous layer of ion-implanted silicon (Si) in the late 1970s [10] and then extended to annealing the lattices of other semiconductors, such as Si [11-23], GaAs [2,15,24,25], InSb [1,3,26-29], and Ge [30,31]. Soon after the discovery of so-called pulsed laser annealing, it was established that such laser annealing is an ultrafast nonthermal melting process [32-35] in which the photoexcited electrons are hot and the ions are still cold (in terms of kinetic energy) because the lattice disordering starts and finishes well before the completion of carrier–lattice thermalization via electron–phonon coupling. Specifically, at sufficiently high levels of photoexcitation, the loss of long-range order inside semiconductor lattices was observed to exist on a sub-picosecond timescale [1], arising from a strong modification of the interatomic potential owing to photoexcitation of a significant amount (10% or more) of electrons from the valence band to the conduction band. This is in sharp contrast to the laser-induced thermal melting in metals, which exists on a timescale of tens of picoseconds due to the required time for electron–lattice equilibration to heat the lattice above the melting point [36,37], followed by liquid nucleation on the surface and spreading out with a liquid front propagating at most at the speed of sound ($1.5 \times 10^3 \, m/s$ for Si) [37-39].

    Pulsed laser annealing in semiconductors has also generated intensive debate over the last four decades about its underlying microscopic mechanisms. It is commonly believed that, on the sub-picosecond timescale, the excitation of a large fraction of electrons from bonding valence bands to anti-bonding conduction bands could weaken the lattice and induce a repulsive interatomic force to quickly disorder the lattice without significantly increasing its thermal energy [1-3]. The debate now focuses on the mechanisms and microscopic picture for such a disorder occurs, whether driven by softened phonon modes with an imaginary frequency (corresponding to a saddle point on the potential energy surface) [14,15,20,28] or by the random velocity of each atom while there is a flat potential energy surface [3,29,40]. The former is termed electron-hole plasma-induced phonon instability theory [11], and the latter is the so-called inertial model [3]. Stampfli and Bennemann [41] utilized a tight-binding theory to simulate the laser pulse-induced melting of Si, Ge, and diamond, claiming to derive a dense electron-hole plasma by populating the single-particle energy levels

according to the Fermi–Dirac distribution with an artificially high electron temperature. This plasma represents the laser pulse excitation of valence electrons, although it usually takes approximately 1 ps for the carrier distribution to reach equilibrium from a Fermi–Dirac distribution [2]. The dense electron hole plasma is found to soften and stabilize the transverse acoustic phonons [14,29] and/or the longitudinal optical phonons [15,28] (with an imaginary frequency), which then drives distortion of the lattice. This is a two-temperature model, assuming the electron and lattice have two different temperatures, but each system is in equilibrium. The direct probe of the atomic structure change using the ultrafast time-resolved X-ray diffraction and the observation of little variation as the laser fluence changes from 50 to 100 mJ/cm$^2$ led Lindenberg et al. to formulate an alternative inertial model [3]. The idea behind the inertial model is that the unbound atoms move at the random velocity they had at the time when the bonds were broken. This model is further supported by the finding of temperature-dependent melting rates because the initial random velocities are set by the thermodynamic temperature [29]. However, this has also been disputed by Zijlstra et al. [28] based on an ab initio DFT prediction that dense electron hole plasma softens only the acoustic phonon rather than all modes, as assumed in the inertial model. Hartley et al. [42] showed the melting is faster than that predicted by the inertial model. The observation of forces acting on the atoms after bond breaking led to an additional Coulomb force model [42,43], where the photoexcitation of a large fraction of valence electrons was suggested to trigger the Coulombic repulsion between ions to immediately disorder the lattice.

However, all these models are difficult to understand because the melting effect varies depending on the laser wavelength. Evidence has accumulated that a shorter melting timescale in Si is due to a longer laser wavelength: 2.03-eV laser pulses give rise to nonthermal melting within 100 fs - 200 fs [12], but a shorter laser wavelength of 3.2-eV offers slower melting, with a time scale of approximately 500 fs [13]. Longer melting time is considered to be the result of a delayed onset of the nonthermal melting event, which is currently interpreted as the time needed for the secondary electron cascade to thermalize the electronic system [21]. Furthermore, all these models suggest that nonthermal melting should develop uniformly at atomic scale for the area under irradiation [29,40]. However, this cannot be resolved experimentally, as the probing techniques of ultrafast electron [4,5,37] and X-ray [1,3] diffraction and ultrafast optical spectroscopy [44,45] all measure averaged results over many unit cells, and thus they are less sensitive to an atomic-scale microscopic melting pattern.

In this paper, we reveal that photoinduced ultrafast nonthermal melting occurs via homogeneous nucleation with randomly distributed local seeds rather than simultaneously breaking all the bonds, as suggested by proposed mechanisms. We use newly developed real-time time-dependent density functional theory (rt-TDDFT) by introducing a Boltzmann factor to restore the detailed balance, which is capable of describing the hot carrier cooling process, [46,47] to perform first-principles simulations of the photoexcitation induced nonthermal melting in Si. Without "ad hoc" hypotheses, our simulations closely reproduce experimental data using both 387-nm (3.2 eV) and 610-nm (2.03 eV) laser pulses. Such unprecedented agreement affirms the reliability of our method and allows us to study the microscopic mechanism behind the laser wavelength-dependent melting process. We further reveal that the nonthermal melting is initiated at atomic sites by lattice vibrations induced by random atomic displacements (nucleation of the liquid phase) and amplified by the spatial localization of photoexcited carriers populated at the band-edge states. This localization leads to dynamic instability of the local lattice and creates a self-trapping center for other carriers, which leads to a positive feedback amplification and causes local melting nucleation. Such atomic-scale nucleation seeds are initially distributed randomly over the laser-irradiated layer. They are followed by rapid growth in size and finally connecting, yielding a complete system nonthermal melting within 200 fs. A sufficient amount of photoexcited hot carriers must be cooled down to the band-edges before participating in the amplification of local lattice distortions via self-trapping. The time needed for the cooling of this hot carrier (rather than electron–lattice equilibration) is the response for the longer melting timescales at shorter laser wavelengths, as observed experimentally for Si [13].

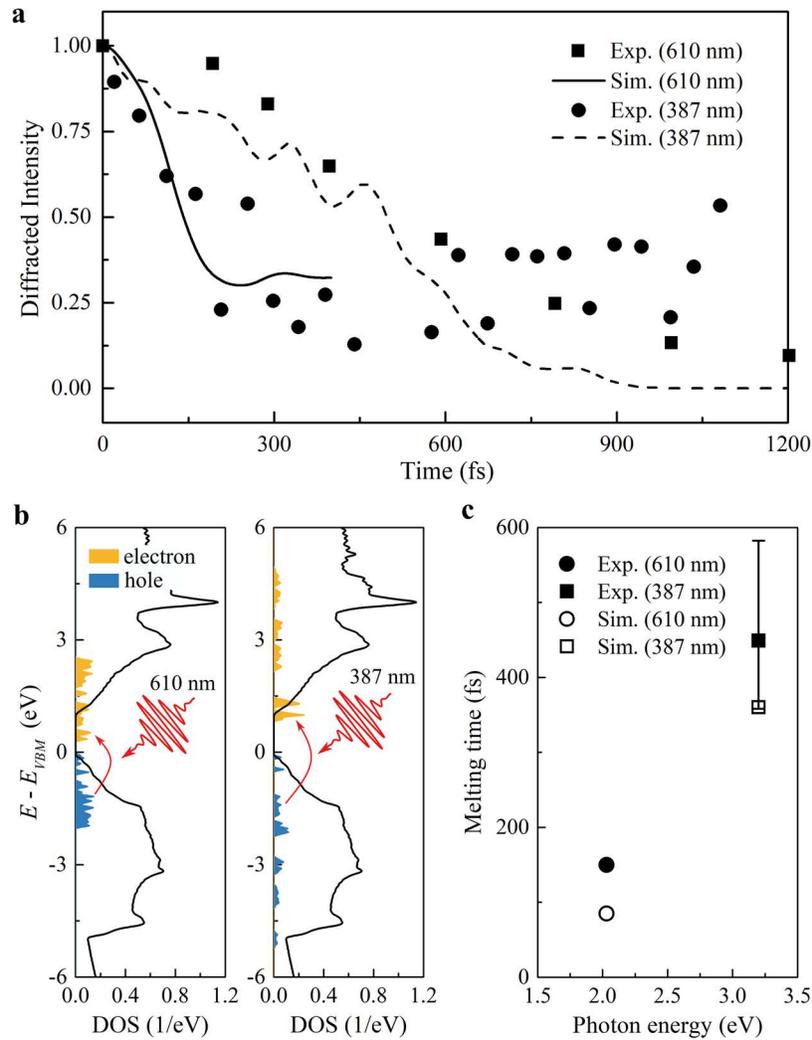

**Figure 1 | Simulation of laser-induced ultrafast nonthermal melting of Si irradiated by two laser pulses with 610-nm and 387-nm wavelengths. a,** The rt-TDDFT simulations predicted dynamic evolution of diffracted intensity following photoexcitation of 11% of valence electrons by 610-nm (solid line) and 387-nm (dashed line) laser wavelength, respectively, in comparison with experimental data (filled square for 610-nm laser [12] and filled circles for 387-nm laser [13]). **b,** Population of energy levels of photoexcited electrons (yellow area) and holes (blue area) at 100 fs following photoexcitation. The black solid line represents the density of states (DOS) of unperturbed bulk Si, where the VBM is set to 0 eV and CBM is at 1.12 eV. **c,** Predicted Si melting (or disordering) times depending on the photon energy (or laser wavelength) is compared with experimental data from Tom *et al.* [12] (610-nm wavelength) and Harb *et al.* [13] (387-nm wavelength).

## Results and Discussion

**Laser wavelength-dependent melting.** To reveal the physics underlying laser wavelength-dependent melting, we investigated the atomic dynamics of the ultrashort laser pulse-induced melting of Si using two different laser wavelengths following experiments [12,13]. This investigation was carried out by performing rt-TDDFT simulations for photoexcited Si at an initial temperature of 300 K. Following the rt-TDDFT simulations, to directly compare the experimental data, we also computed the X-ray diffraction intensity $I(t)$ based on the Debye–Waller formula [3,13,20], $I(t) = \exp[-Q^2 <u^2(t)>/3]$, where $Q$ is the reciprocal lattice vector corresponding to the X-ray reflection peak and $u^2(t)$ is the square of the root-mean-square displacement (RMSD) averaged over all atoms. Figure 1a shows that the simulated $I(t)$ is in excellent agreement with the experimental data for photoexcitation at both laser wavelengths, showing that the laser wavelength has a substantial influence on the melting process. The quantitative agreement with the experiment affirms the accuracy and reliability of our simulations. We indeed reveal a nonintuitive phenomenon that the lower-energy photons cause a much faster nonthermal melting than the high-energy photons (Fig. 1a), although the 387-nm laser pulse deposits 1.6 times more energy in the system than the 610-nm laser pulses, both exciting the same amount (11%) of valence electrons into the conduction bands. The predicted nonthermal melting times agree with experiments within the experimental error bars and demonstrate that the melting rate (1/time) is inversely proportional to the photon energy, as shown in Fig. 1c. The different melting processes can be traced back to the distinct distribution of photoexcited electrons and holes within the conduction and valence bands, as shown in Fig. 1b. We can see that the 610-nm laser pulse excites the valence electrons mostly from the top part of the valence band to populate the bottom part of the conduction band, whereas the 387-nm laser pulse promotes the electrons from the deeper part of the valence band into the higher energy levels in the conduction band, spanning a wide energy range of the valence and conduction bands.

**The emergence of nucleation seeds.** We now examine the detailed atomic dynamics of Si following irradiation with a 610-nm laser pulse exciting 11% of valence electrons according to experiments [12]. Following common practice in the literature [18], we adopt RMSD to characterize the degree of the loss of long-range atomic disorder (or melting). The usual criterion used to determine melting is the Lindemann criterion at 15% [18,20], which is RMSD = 0.35 Å as the Si equilibrium bond length is 2.35 Å, as indicated by the dashed line in Fig. 2c. We find that the value of RMSD increases monotonically during the first 200 fs following photoexcitation: it reaches the Lindemann criterion (0.35 Å) at approximately 100 fs and continues to grow to 0.8 Å at 200 fs, as shown in Fig. 2a. It is interesting to see in Fig. 2b that throughout the whole simulation, the lattice remains relatively cool with a temperature of approximately 450 K, which is far below the melting temperature of Si (1680 K) [16]. This demonstrates a nonthermal melting process.

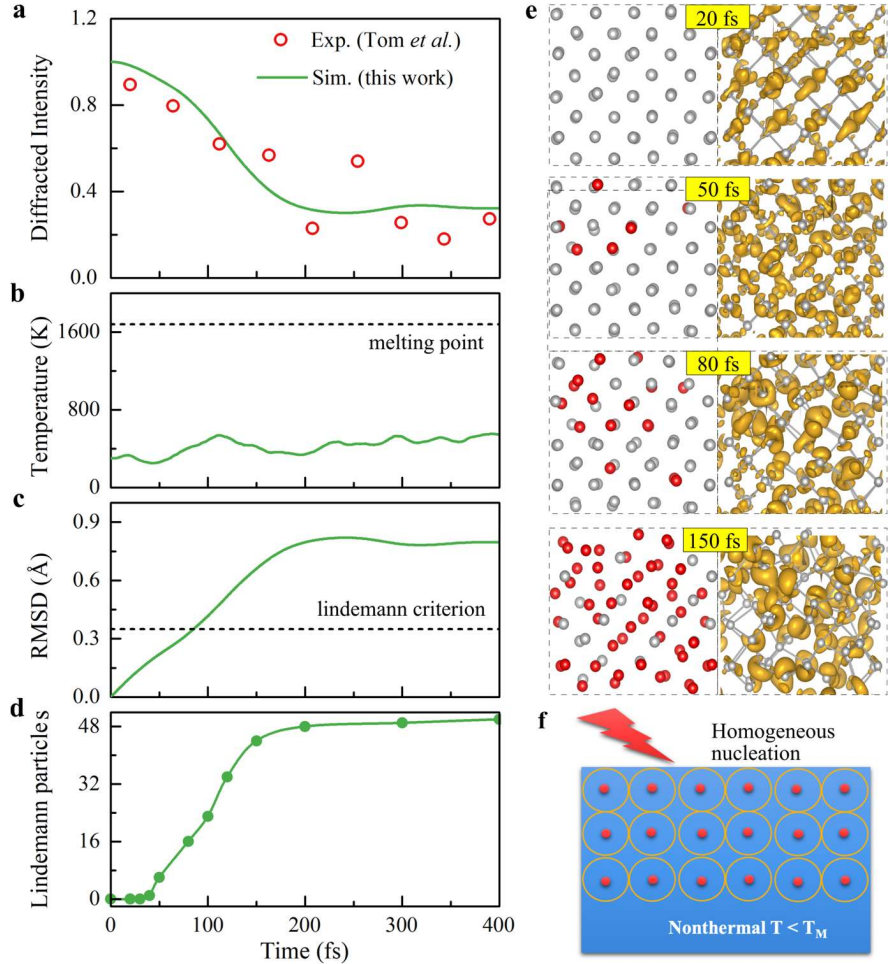

**Figure 2 | Atomic dynamics of laser-induced ultrafast melting of Si irradiated with a 610-nm, 100-fs laser pulse. a,** Time-dependent diffraction intensity obtained directly from the evolution of atomic positions in the rt-TDDFT simulation (green line) in comparison with the experimental data (red circles). [12] **b,** Lattice temperature as a function of time following the photoexcitation. The dashed line marks the Si melting temperature $T_M$ = 1680 K. **c,** Simulation predicted root-mean-square displacement (RMSD) of atoms as a function of time following the photoexcitation. The dashed line indicates the Lindemann criterion, which is $R_c$ = 0.35 Å for 15% of equilibrium Si-Si bond length 2.35 Å).[18] **d,** Number of Lindemann particles following the photoexcitation. **e,** Snapshots of atomic displacements (viewed along the [001] direction)

with time at 0, 50, 80, and 150 fs, respectively, following the photoexcitation. The red atoms indicate the Lindemann particles with displacements $R_i(t) - R_i(0) > 0.35$ Å, which represents the molten atoms. The corresponding real-space distributions of photoexcited electrons (yellow iso-surface) are in the right panel. **f,** Schematically illustrated the homogeneous nucleation of the laser-induced ultrafast nonthermal melting. The red points represent the randomly distributed nucleation seeds corresponding to clusters of Lindemann particles).

To reveal the microscopic details of the nonthermal melting, we plot a series of snapshots with time (0, 50, 80, and 100 fs following photoexcitation) for atomic structures of photoexcited Si in Fig. 2e. We render atoms with displacements exceeding 0.35 Å (so-called "Lindemann particles") in red in Fig. 2e. These Lindemann particles consisting of single atoms emerge as early as 50 fs (Fig. 2d), although the melting deduced from the diffraction intensity starts only at 100 fs and completes (indicated by the minimum in the diffraction intensity) at 200 fs. Afterward, these Lindemann particles develop into a cluster, which grows rapidly in size, as illustrated in the 80 fs and 150 fs snapshots. Consequently, the molten atoms in each supercell spread outwards and then connect with its nearest-neighbor images in the supercell calculation, yielding ultrafast nonthermal melting throughout the whole system within 200 fs. In short, these Lindemann particles occur randomly and statistically inside the bulk crystal and are interpreted as the seeds for melting nucleation, which eventually causes melting of the whole system, as schematically illustrated in Fig. 2f (also see Supplementary Fig. 1).

We thus reveal that nonthermal melting occurs via a homogenous nucleation process with randomly distributed local seeds, rather than all at once throughout the Si crystal in an atomic-scale uniform fashion, as suggested by the inertial model [3] and phonon instability theory [14,15] (due to the simultaneous breaking of all bonds caused by the excitation of high-density electron-hole plasma). It has been established that homogeneous nucleation is a mechanism for the rapid thermal melting of superheated metals within several ps [36,37,39,48,49]. It was predicted that in superheated metals, the vibrational and mechanical instabilities of the lattice occur simultaneously but only locally, leading to the formation of destabilized clusters inside the bulk [50]. The melting rate is mostly governed by electron–lattice equilibration (approximately several ps) to heat the atoms in the metal to a superheated state. In our semiconductor case, electron–lattice equilibrium is not realized at the nonthermal melting time, and our melting is not caused by superheating. While the nucleation phenomenon appears to be the same, the atomistic mechanisms causing such local nucleations are different.

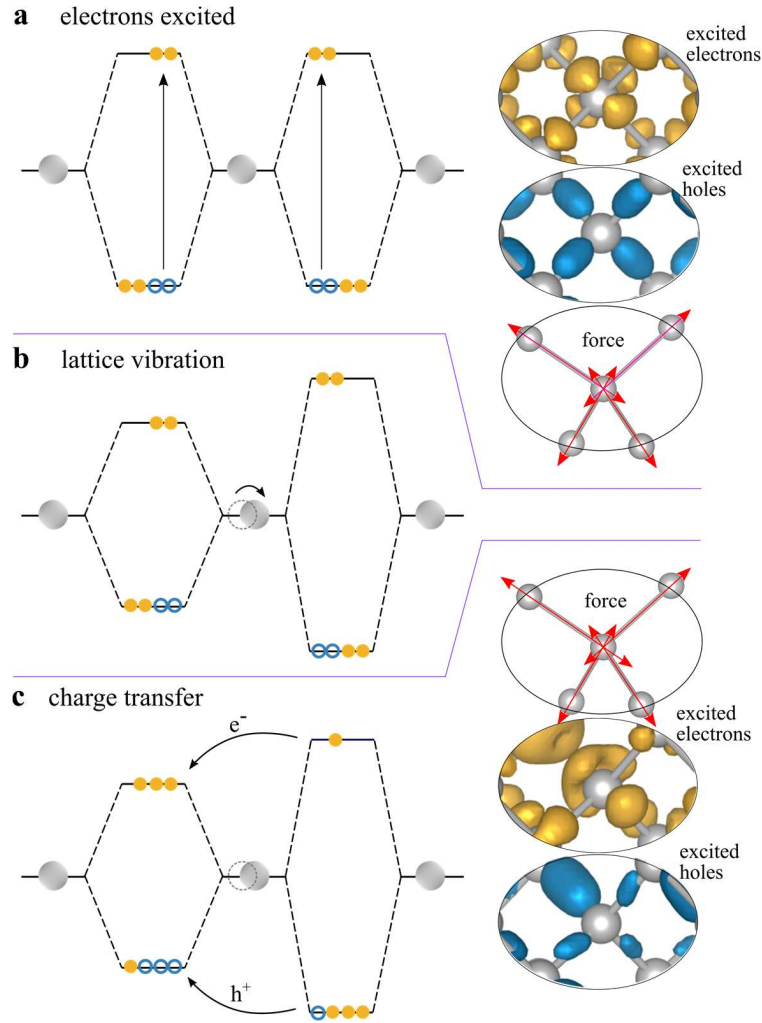

**Figure 3 | Emergence of the self-trapping center due to electron–lattice coupling-induced self-amplification. a,** The incident photons promote valence electrons from the filled bonding states in the valence band to the empty anti-bonding states in the conduction band. The photoexcited electrons and holes are uniformly distributed among atom bonds, generating interatomic forces with an equivalent magnitude. **b,** The lattice vibration-induced random movements of atoms cause local distortions in the lattice, which in turn yields band-edge fluctuations. **c,** The tensile distorted region acts as a trapping center for excited electrons and holes and induces carrier localization. The carrier localization, in turn, amplifies the initial thermal random distortion. Subsequently, this process is an unstable self-trapping process and is slightly similar to the polaron to break the translational symmetry [51]. Insets to (**a**) and (**c**) represent the real-space distribution of excited electrons and holes, as well as charge population, generated interatomic forces immediately after photoexcitation and at 200 fs, respectively.

**The microscopic mechanism underlying the birth of nucleation seeds.** We delineate the factors that break the spatial translational symmetry and locally cause atomic disorders as seeds of homogeneous nucleation. The Lindemann particles emerge as early as 50 fs, which is in the middle of the laser pulse irradiation since the laser pulse duration is approximately 100 fs. This means the melting starts even when the laser has only excited 9% of valence electrons, which is less than the threshold predicted under the electron–hole plasma model [14,15,41]. In our above simulations, the system is in equilibrium at room temperature before the photoexcitation. There are all kinds of phonons in Si, such as the acoustic and optical phonons at the zone center belonging to symmetry-breaking $T_{1u}$ and $T_{2g}$ group representations, respectively, rather than the $A_{1g}$ representation that is the only phonon mode for the lattice to vibrate while preserving symmetry. The thermal random movements of atoms of the phonon modes and the associated breaking of translational symmetry can obscure the analysis of electronic properties and interatomic forces, hindering the disclosure of origin of the birth of nucleation seeds. For instance, the real-space distribution of photoexcited electrons for a snapshot at 20 fs exhibits nonuniformity, as shown in Fig. 2e. Such randomness can obscure the true reason for the nucleation. To suppress the effects of such thermal random movements, we repeated the simulation for a 610-nm laser pulse at a low initial temperature of 1 K (Supplementary Fig. S2). At the initial stage following the photoexcitation, the photoexcited electrons and holes shown in Fig. 3a are uniformly and coherently distributed over the whole lattice. The photoexcitation induces the depletion of the bonding states and population of the anti-bonding states, not disturbed by any significant randomness at the initial time. The additional interatomic forces induced by the photoexcitation are uniform for all the atoms, resulting in a zero net force on each Si atom (schematically shown in Fig. 3a). However, with time, carrier localizations occur unexpectedly in both excited electrons and holes, as shown in Fig. 3c. This random localization is initiated by the original small atomic vibrations at 1 K. Such localization traps the excited carriers in some local regions, which causes local atomic displacement, which in term traps more carriers, thus forming a positive instability loop. This leads to the formation of melting seeds. It is worth mentioning that such carrier localization, which is critical to our model, does not play any role in the inertial model, phonon softening model, and Coulomb force models. We analyze below why such local carrier trapping is dynamically unstable with a self-trapping and self-amplification effect.

One can simplify the electronic structure of bulk Si as the bonding and anti-bonding states of Si covalent bonds, as shown in Fig. 3a-c. At the ground state, the bonding states are fully occupied by valence electrons leaving the anti-bonding states empty with a 1.1 eV bandgap between them. Under laser irradiation, incident photons excite electrons from the bonding states (valence band) to the anti-bonding states (conduction band) leaving behind voids (holes) in the bonding states (Fig. 3a). The electron (hole) population of the anti-bonding (bonding) states raise the free energy of the excited system, which can lower its energy by lowering (raising) the energy level of the anti-bonding (bonding) states by elongating the Si-Si bond length. As a result, the charge population generates a stretching force on each bond. However, if everything is uniform and coherent atomically, the starching forces at all the bonds exert a zero net force on each atom. However, if there are small fluctuations (as provided by the 1K initial randomness), the bond with a slightly longer length lowers (raises) the anti-bonding (bonding) state more than its neighbors. This further traps more electrons (holes) at the anti-bonding (bonding) state, causing a larger stretching force on the end atoms. This causes an imbalance in the net force at each atom, thus further stretching the bond length. This causes an unstable amplification mechanism. That is exactly what is shown in Fig. 3c. We thus propose an excited carrier localization instability with a carrier self-trapping picture. This picture is different from both the inertial model and phonon softening model. Because of the carrier localization, the density of excited electrons and holes in the local region easily exceeds the required density for phonon softening, although the corresponding overall charge density is still lower than that the requirement in the phonon softening theory [14,15,41]. This explains why the system at 50 fs with only 9% overall valence electron excitation already has some local melting at the nucleation sites.

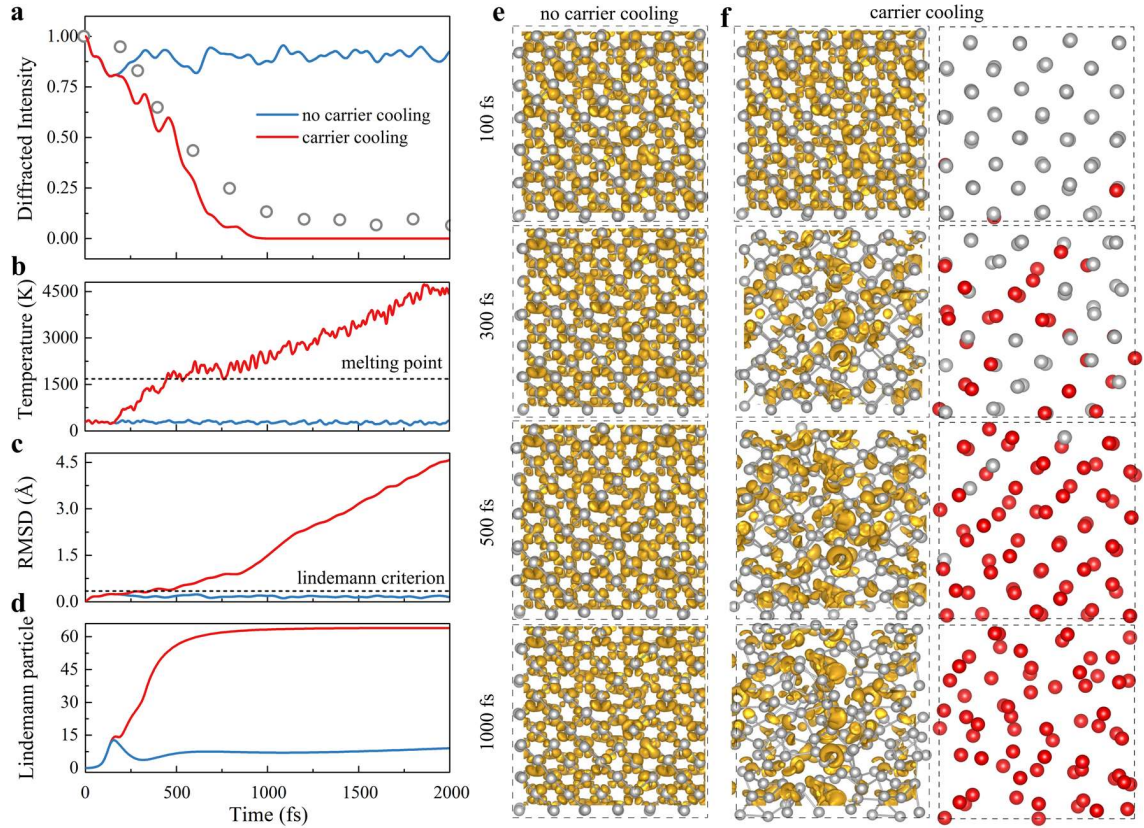

**Figure 4 | Atomic dynamics of laser-induced ultrafast melting of Si irradiated with a 387-nm, 150-fs laser pulse. a,** The rt-TDDFT simulation predicted electron diffracted intensity of Si (220) Bragg peaks following the photoexcitation by a 387-nm 150-fs pump pulse in comparison with experimental data[13]. Here, we carry out two rt-TDDFT simulations: one takes into account the hole carrier cooling (red lines) and another has no carrier cooling, as in conventional rt-TDDFT simulations (blue lines). **b,** Evolution of crystal temperature following the photoexcitation. **c,** Evolution of atomic RMSD following the photoexcitation. **d,** Evolution of the number of Lindemann particles. **e, f,** Snapshoots of real-space distributions of excited electrons (yellow iso-surface) and atomic displacements with times at 100, 300, 500, and 1000 fs following the photoexcitation for (e) without and (f) with taking into account carrier cooling, respectively. The red atoms indicate the Lindemann particles with displacements $R_i(t) - R_i(0) > 0.35$ Å.

**Role of hot carrier cooling.** The reason why the melting is slower when a higher frequency laser is used remains to be explained. The increased melting time of shorter laser wavelengths has caused controversy in some studies [12,13,22]. As we discussed above, ultrafast nonthermal melting mainly occurs due to the self-trapping of excited electrons and holes at the band-edge states, which weakens the Si-Si bonds at atomic sites. Hot carriers populated at higher energy levels have to relax to the band-edge states first before participating in the self-trapping process. Figure 1c shows that the 610-nm laser pulse excited electrons and holes are much closer to the band edges than the 387-nm laser pulse case. Higher energy states have more delocalization and thus have weak bonding and anti-bonding characteristics than the band-edge states, thus contributing less to Si-Si bond weakening. To demonstrate this further, we used a constrained DFT to calculate the potential energy surfaces (PESs) along the Si-Si bond stretching using carrier population, which is the same as in the cases of 610-nm and 387-nm laser wavelengths (Supplementary Fig. S3). This shows that the carrier population according to the case of a 610-nm laser wavelength can remarkably modify the PES to weaken the lattice, whereas the carrier population in the case of a 387-nm laser wavelength hardly has any effects in modifying the PES. This illustrates that ultrafast nonthermal melting is not enhanced by deposing more energy into the electronic subsystem by increasing the single-photon energy; rather, the occupations of the special states (band-edge states) are more responsible. Furthermore, we also carried out conventional rt-TDDFT simulations without the use of the Boltzmann factor technique, which cannot describe the carrier cooling process due to the lack of a detailed balance [52]. In such conventional rt-TDDFT simulations, 387-nm laser-excited electrons and holes always occupy higher energy levels. No melting occurs during the 2 ps simulation time, and the excited carrier charge density distributes evenly in space throughout the whole Si crystal during all simulation times (Fig. 4e).

Of course, experimental measurements have shown that under 387-nm laser pulse irradiation, nonthermal melting does occur, albeit at a longer timescale of 500 fs. In reality, excited hot carriers relax to lower energy levels through the emission of phonons and transfer excess energy into the lattice (hot carrier cooling) [2,53]. Our newly developed rt-TDDFT algorithm with the Boltzmann factor (Supplementary Note 3) can describe this hot carrier cooling properly. In the Boltzmann-TDDFT simulation, our results have indeed shown a longer timescale (~ 500 fs) of nonthermal melting for the 387-nm laser pulse, in excellent agreement with experimental observations [13], as shown in Fig. 1a and Fig. 4d. It is interesting to note that the first Lindemann particle appears as early as 80 fs, implying that it originates from the localization of a small number of photoexcited carriers inside the locally distorted atomic sites due to the self-trapping and self-amplification process, as we discussed above. Time is needed for more hot carriers to relax to band-edge states to participate in the growth of the nucleation seeds. This is confirmed by the fact that the lattice is heated above the melting temperature within 500 fs, as shown in Fig. 4b. Therefore, we conclude that the longer melting start time in the 387-nm laser case is due to hot carrier cooling. Overall, we show that hot carrier cooling is essential. Not only is it needed to describe the self-trapping and amplification process near the band-edge states, but it is also needed to describe the cooling from the higher energy excitation to the band-edge states.

**Conclusion**

In summary, based on rt-TDDFT simulations using a newly developed Boltzmann method, we have proposed homogeneous nucleation with randomly distributed seeds arising from local instability caused by carrier self-trapping and amplifications as the microscopic mechanism for the laser-induced ultrafast nonthermal melting of semiconductors. Homogeneous nucleation has also been proposed before as a mechanism for the rapid melting of superheated metals [37,48,50] with a time scale above several ps. In such metallic cases, the local instability is caused by the thermal vibration of the lattice. In contrast, for semiconductors, the local instability is driven mostly by the photoexcited carriers via a self-trapping and self-amplification process. This is only possible when there is a bandgap in the material (semiconductor) with bonding and anti-bonding states at the band edges, which is fundamentally different from the metallic case. Due to this local dynamic instability, any initial small random displacements induced by phonon vibrations can be amplified and followed by charge transfer of the photoexcited carriers, which in turn weakens the local bond and attracts more carriers. This amplifies the initial thermal randomness, yielding local nucleation seeds for nonthermal melting. Because a sufficient amount of photoexcited hot carriers must be cooled down to the band edges before participating in the self-amplification of local lattice distortions, the time needed to cool this portion of hot carriers (rather than electron–lattice equilibration) is the response for the longer melting timescales at shorter laser wavelengths. We believe our results offer a comprehensive and detailed picture for nonthermal melting in semiconductors and provide fresh insights into photoinduced dynamics.

**Methods**

We carry out the real-time time-dependent density functional theory (rt-TDDFT) simulations [54] based on norm-conserving pseudopotentials (NCPP) [55] and the Perdew–Burke–Ernzerhof (PBE) functional within a density functional theory (DFT) framework with a plane wave nonlocal pseudopotential Hamiltonian, which is implemented in the code PWmat [56]. The wave functions are expanded on a plane-wave basis with an energy cutoff of 50 Ry. In rt-TDDFT simulations, we use a 64-atom supercell for Si, and the Γ point is used to sample the Brillouin zone. The time step is set to 0.1 fs in our dynamic simulations. The laser pulse has a wavelength λ = 610 nm, duration 2σ = 25 fs, and photon energy ω = 2.03 eV, which is consistent with the parameters from another experiment [12]. Another laser pulse has a wavelength λ = 387 nm, duration 2σ = 25 fs, and photon energy ω = 3.2 eV, which is consistent with the parameters from another experiment [13] (Supplementary Note 1). The new Boltzmann factor algorithm for the rt-TDDFT can be found in Ref. 46 and 47 and Supplementary Note 3.

## Data availability

The data that support the findings of this study are available from the corresponding authors upon reasonable request.

## Code availability

The rt-TDDFT CODE has been integrated into the PWmat package. The PWmat software can also be accessed directly from http://www.pwmat.com.

1. Rousse, A. *et al.* Non-thermal melting in semiconductors measured at femtosecond resolution. *Nature* **410**, 65-68, doi:10.1038/35065045 (2001).
2. Sundaram, S. K. & Mazur, E. Inducing and probing non-thermal transitions in semiconductors using femtosecond laser pulses. *Nat Mater* **1**, 217-224, doi:10.1038/nmat767 (2002).
3. Lindenberg, A. M. *et al.* Atomic-scale visualization of inertial dynamics. *Science* **308**, 392-395, doi:10.1126/science.1107996 (2005).
4. Eichberger, M. *et al.* Snapshots of cooperative atomic motions in the optical suppression of charge density waves. *Nature* **468**, 799-802, doi:10.1038/nature09539 (2010).
5. Sciaini, G. & Miller, R. J. D. Femtosecond electron diffraction: heralding the era of atomically resolved dynamics. *Reports on Progress in Physics* **74**, doi:10.1088/0034-4885/74/9/096101 (2011).
6. Frigge, T. *et al.* Optically excited structural transition in atomic wires on surfaces at the quantum limit. *Nature* **544**, 207-211, doi:10.1038/nature21432 (2017).
7. Seddon, E. A. *et al.* Short-wavelength free-electron laser sources and science: a review. *Rep Prog Phys* **80**, 115901, doi:10.1088/1361-6633/aa7cca (2017).
8. Kogar, A. *et al.* Light-induced charge density wave in LaTe3. *Nature Physics* **16**, 159-163, doi:10.1038/s41567-019-0705-3 (2019).
9. Horstmann, J. G. *et al.* Coherent control of a surface structural phase transition. *Nature* **583**, 232-236, doi:10.1038/s41586-020-2440-4 (2020).
10. Young, R. T. *et al.* Laser annealing of boron‐implanted silicon. *Applied Physics Letters* **32**, 139-141, doi:10.1063/1.89959 (1978).
11. Combescot, M. & Bok, J. Instability of the Electron-Hole Plasma in Silicon. *Physical Review Letters* **48**, 1413-1416, doi:10.1103/PhysRevLett.48.1413 (1982).

12   Tom, H. W., Aumiller, G. D. & Brito-Cruz, C. H. Time-resolved study of laser-induced disorder of Si surfaces. *Phys Rev Lett* **60**, 1438-1441, doi:10.1103/PhysRevLett.60.1438 (1988).
13   Harb, M. *et al.* Electronically driven structure changes of Si captured by femtosecond electron diffraction. *Phys Rev Lett* **100**, 155504, doi:10.1103/PhysRevLett.100.155504 (2008).
14   Stampfli, P. & Bennemann, K. H. Dynamical theory of the laser-induced lattice instability of silicon. *Phys Rev B Condens Matter* **46**, 10686-10692, doi:10.1103/physrevb.46.10686 (1992).
15   Stampfli, P. & Bennemann, K. H. Time dependence of the laser-induced femtosecond lattice instability of Si and GaAs: Role of longitudinal optical distortions. *Phys Rev B Condens Matter* **49**, 7299-7305, doi:10.1103/physrevb.49.7299 (1994).
16   Silvestrelli, P. L., Alavi, A., Parrinello, M. & Frenkel, D. Ab initio Molecular Dynamics Simulation of Laser Melting of Silicon. *Phys Rev Lett* **77**, 3149-3152, doi:10.1103/PhysRevLett.77.3149 (1996).
17   Recoules, V., Clerouin, J., Zerah, G., Anglade, P. M. & Mazevet, S. Effect of intense laser irradiation on the lattice stability of semiconductors and metals. *Phys Rev Lett* **96**, 055503, doi:10.1103/PhysRevLett.96.055503 (2006).
18   Zijlstra, E. S., Kalitsov, A., Zier, T. & Garcia, M. E. Squeezed Thermal Phonons Precurse Nonthermal Melting of Silicon as a Function of Fluence. *Physical Review X* **3**, 011005 doi:10.1103/PhysRevX.3.011005 (2013).
19   Medvedev, N., Li, Z. & Ziaja, B. Thermal and nonthermal melting of silicon under femtosecond x-ray irradiation. *Physical Review B* **91**, 054113 doi:10.1103/PhysRevB.91.054113 (2015).
20   Lian, C., Zhang, S. B. & Meng, S. Ab initio evidence for nonthermal characteristics in ultrafast laser melting. *Physical Review B* **94**, 184310, doi:10.1103/PhysRevB.94.184310 (2016).
21   Pardini, T. *et al.* Delayed Onset of Nonthermal Melting in Single-Crystal Silicon Pumped with Hard X Rays. *Phys Rev Lett* **120**, 265701, doi:10.1103/PhysRevLett.120.265701 (2018).
22   Darkins, R., Ma, P.-W., Murphy, S. T. & Duffy, D. M. Simulating electronically driven structural changes in silicon with two-temperature molecular dynamics. *Physical Review B* **98**, 024304, doi:10.1103/PhysRevB.98.024304 (2018).
23   Medvedev, N., Kopecky, M., Chalupsky, J. & Juha, L. Femtosecond x-ray diffraction can discern nonthermal from thermal melting. *Physical Review B* **99**, 100303(R), doi:10.1103/PhysRevB.99.100303 (2019).
24   Saeta, P., Wang, J., Siegal, Y., Bloembergen, N. & Mazur, E. Ultrafast electronic disordering during femtosecond laser melting of GaAs. *Phys Rev Lett* **67**, 1023-1026, doi:10.1103/PhysRevLett.67.1023 (1991).
25   Graves, J. S. & Allen, R. E. Response of GaAs to fast intense laser pulses. *Physical Review B* **58**, 13627-13633, doi:DOI 10.1103/PhysRevB.58.13627 (1998).
26   Gaffney, K. J. *et al.* Observation of structural anisotropy and the onset of liquidlike motion during the nonthermal melting of InSb. *Phys Rev Lett* **95**, 125701, doi:10.1103/PhysRevLett.95.125701 (2005).
27   Lindenberg, A. M. *et al.* X-ray diffuse scattering measurements of nucleation dynamics at femtosecond resolution. *Phys Rev Lett* **100**, 135502, doi:10.1103/PhysRevLett.100.135502 (2008).
28   Zijlstra, E. S., Walkenhorst, J. & Garcia, M. E. Anharmonic noninertial lattice dynamics during ultrafast nonthermal melting of InSb. *Phys Rev Lett* **101**, 135701, doi:10.1103/PhysRevLett.101.135701 (2008).
29   Wang, X. *et al.* Role of Thermal Equilibrium Dynamics in Atomic Motion during Nonthermal Laser-Induced Melting. *Phys Rev Lett* **124**, 105701, doi:10.1103/PhysRevLett.124.105701 (2020).
30   Siders, C. W. *et al.* Detection of nonthermal melting by ultrafast X-ray diffraction. *Science* **286**, 1340-1342, doi:10.1126/science.286.5443.1340 (1999).
31   Sokolowski-Tinten, K. *et al.* Femtosecond x-ray measurement of ultrafast melting and large acoustic transients. *Phys Rev Lett* **87**, 225701, doi:10.1103/PhysRevLett.87.225701 (2001).

32  Van Vechten, J. A., Tsu, R., Saris, F. W. & Hoonhout, D. Reasons to believe pulsed laser annealing of Si does not involve simple thermal melting. *Physics Letters A* **74**, 417-421, doi:10.1016/0375-9601(79)90241-x (1979).

33  Van Vechten, J. A., Tsu, R. & Saris, F. W. Nonthermal pulsed laser annealing of Si; plasma annealing. *Physics Letters A* **74**, 422-426, doi:10.1016/0375-9601(79)90242-1 (1979).

34  Shank, C. V., Yen, R. & Hirlimann, C. Time-Resolved Reflectivity Measurements of Femtosecond-Optical-Pulse-Induced Phase Transitions in Silicon. *Physical Review Letters* **50**, 454-457, doi:10.1103/PhysRevLett.50.454 (1983).

35  Shank, C. V., Yen, R. & Hirlimann, C. Femtosecond-Time-Resolved Surface Structural Dynamics of Optically Excited Silicon. *Physical Review Letters* **51**, 900-902, doi:10.1103/PhysRevLett.51.900 (1983).

36  Rethfeld, B., Sokolowski-Tinten, K., von der Linde, D. & Anisimov, S. I. Ultrafast thermal melting of laser-excited solids by homogeneous nucleation. *Physical Review B* **65**, doi:10.1103/PhysRevB.65.092103 (2002).

37  Mo, M. Z. *et al.* Heterogeneous to homogeneous melting transition visualized with ultrafast electron diffraction. *Science* **360**, 1451-1455, doi:10.1126/science.aar2058 (2018).

38  Phillpot, S. R., Yip, S. & Wolf, D. How Do Crystals Melt? *Computers in Physics* **3**, doi:10.1063/1.4822877 (1989).

39  Lin, Z. & Zhigilei, L. V. Time-resolved diffraction profiles and atomic dynamics in short-pulse laser-induced structural transformations: Molecular dynamics study. *Physical Review B* **73**, doi:10.1103/PhysRevB.73.184113 (2006).

40  Inoue, I. *et al.* Atomic-Scale Visualization of Ultrafast Bond Breaking in X-Ray-Excited Diamond. *Phys Rev Lett* **126**, 117403, doi:10.1103/PhysRevLett.126.117403 (2021).

41  Stampfli, P. & Bennemann, K. H. Theory for the instability of the diamond structure of Si, Ge, and C induced by a dense electron-hole plasma. *Phys Rev B Condens Matter* **42**, 7163-7173, doi:10.1103/physrevb.42.7163 (1990).

42  Hartley, N. J. *et al.* Using Diffuse Scattering to Observe X-Ray-Driven Nonthermal Melting. *Phys Rev Lett* **126**, 015703, doi:10.1103/PhysRevLett.126.015703 (2021).

43  Medvedev, N., Jeschke, H. O. & Ziaja, B. Nonthermal phase transitions in semiconductors induced by a femtosecond extreme ultraviolet laser pulse. *New Journal of Physics* **15**, doi:10.1088/1367-2630/15/1/015016 (2013).

44  Rohwer, T. *et al.* Collapse of long-range charge order tracked by time-resolved photoemission at high momenta. *Nature* **471**, 490-493, doi:10.1038/nature09829 (2011).

45  Nicholson, C. W. *et al.* Beyond the molecular movie: Dynamics of bands and bonds during a photoinduced phase transition. *Science* **362**, 821-825, doi:10.1126/science.aar4183 (2018).

46  Wang, L. W. Natural Orbital Branching Scheme for Time-Dependent Density Functional Theory Nonadiabatic Simulations. *J Phys Chem A* **124**, 9075-9087, doi:10.1021/acs.jpca.0c06367 (2020).

47  Liu, W.-H. *et al.* Algorithm advances and applications of time-dependent first-principles simulations for ultrafast dynamics. *WIREs Computational Molecular Science* **n/a**, e1577, doi:https://doi.org/10.1002/wcms.1577 (2021).

48  Lu, K. & Li, Y. Homogeneous nucleation catastrophe as a kinetic stability limit for superheated crystal. *Physical Review Letters* **80**, 4474-4477, doi:DOI 10.1103/PhysRevLett.80.4474 (1998).

49  Mazevet, S., Clerouin, J., Recoules, V., Anglade, P. M. & Zerah, G. Ab-initio simulations of the optical properties of warm dense gold. *Phys Rev Lett* **95**, 085002, doi:10.1103/PhysRevLett.95.085002 (2005).

50  Jin, Z. H., Gumbsch, P., Lu, K. & Ma, E. Melting mechanisms at the limit of superheating. *Phys Rev Lett* **87**, 055703, doi:10.1103/PhysRevLett.87.055703 (2001).

51       Zhang, L. *et al.* Dynamics of Photoexcited Small Polarons in Transition-Metal Oxides. *J Phys Chem Lett* **12**, 2191-2198, doi:10.1021/acs.jpclett.1c00003 (2021).
52       Liu, W.-H., Luo, J.-W., Li, S.-S. & Wang, L.-W. The critical role of hot carrier cooling in optically excited structural transitions. *npj Computational Materials* **7**, 117, doi:10.1038/s41524-021-00582-w (2021).
53       YU, P., Cardona, Manuel. *Fundamentals of Semiconductors*.  (2010).
54       Wang, Z., Li, S. S. & Wang, L. W. Efficient real-time time-dependent density functional theory method and its application to a collision of an ion with a 2D material. *Phys Rev Lett* **114**, 063004, doi:10.1103/PhysRevLett.114.063004 (2015).
55       Hamann, D. R. Optimized norm-conserving Vanderbilt pseudopotentials. *Physical Review B* **88**, 085117, doi:ARTN 085117
10.1103/PhysRevB.88.085117 (2013).
56       Jia, W. L. *et al.* The analysis of a plane wave pseudopotential density functional theory code on a GPU machine. *Computer Physics Communications* **184**, 9-18, doi:10.1016/j.cpc.2012.08.002 (2013).
57       Ren, J., Vukmirović, N. & Wang, L.-W. Nonadiabatic molecular dynamics simulation for carrier transport in a pentathiophene butyric acid monolayer. *Physical Review B* **87**, doi:10.1103/PhysRevB.87.205117 (2013).
58       Parandekar, P. V. & Tully, J. C. Detailed Balance in Ehrenfest Mixed Quantum-Classical Dynamics. *J Chem Theory Comput* **2**, 229-235, doi:10.1021/ct050213k (2006).
59       Parandekar, P. V. & Tully, J. C. Mixed quantum-classical equilibrium. *J Chem Phys* **122**, 094102, doi:10.1063/1.1856460 (2005).

**Acknowledgments**

The work in China was supported by the Key Research Program of Frontier Sciences，CAS under Grant No. ZDBS-LY-JSC019, CAS Project for Young Scientists in Basic Research under Grant No. YSBR-026, the Strategic Priority Research Program of the Chinese Academy of Sciences under Grant No. XDB43020000, and the National Natural Science Foundation of China (NSFC) under Grant Nos. 11925407 and 61927901. L.W. W was supported by the Director, Office of Science, the Office of Basic Energy Sciences (BES), Materials Sciences and Engineering (MSE) Division of the U.S. Department of Energy (DOE) through the theory of material (KC2301) program under Contract No. DEAC02-05CH11231.

**Author Contributions**

W.L. performed the TDDFT simulations and prepared the figures. J.L. and L.W. proposed the research project, established the project direction, and conducted the analysis, discussion, and writing of the paper with input from W.L. S.L. provided the project infrastructure, and supervised W.L.'s study.

**Competing interests**

The authors declare no competing interests.

Supplemental Information for **"The seeds and homogeneous nucleation of photoinduced nonthermal melting in semiconductors due to self-amplified local dynamic instability"**

Wen-Hao Liu[1,2], Jun-Wei Luo[1,2,3]*, Shu-Shen Li[1,2,3], and Lin-Wang Wang[4]*

[1]State Key Laboratory of Superlattices and Microstructures, Institute of Semiconductors, Chinese Academy of Sciences, Beijing 100083, China

[2]Center of Materials Science and Optoelectronics Engineering, University of Chinese Academy of Sciences, Beijing 100049, China

[3]Beijing Academy of Quantum Information Sciences, Beijing 100193, China

[4]Materials Science Division, Lawrence Berkeley National Laboratory, Berkeley, California 94720, United States

*Email: jwluo@semi.ac.cn; lwwang@lbl.gov

# Note 1: rt-TDDFT methods for photoexcited dynamic simulations

In this rt-TDDFT algorithm, the time-dependent wave functions, $\psi_j(t)$, are expanded by the adiabatic eigenstates, $\phi_i(t)$:

$$\psi_j(t) = \sum_i C_{j,i}(t)\phi_i(t) \tag{1}$$

and

$$H(t)\phi_i(t) = \epsilon_i(t)\phi_i(t) \tag{2}$$

Here, $H(t) \equiv H(t, R(t), \rho(t))$, $R(t)$ represents the nuclear positions, and $\rho(t)$ represents the charge density. By using equation (1), the evolution of the wave functions $\psi_j(t)$ is changed to the evolution of the coefficient $C_{j,i}(t)$. In equation (2), a linear-time-dependent Hamiltonian (LTDH) is applied to represent the time dependence of the Hamiltonian within a time step $[t_1, t_1 + \Delta t]$: For any $t \in [t_1, t_1 + \Delta t]$

$$H(t) = H(t_1) + \frac{t - t_1}{\Delta t}[H(t_1 + \Delta t) - H(t_1)] \tag{3}$$

Thus, we can obtain a much larger time step (0.1 fs - 0.2 fs) than the conventional real-time TDDFT (sub-attosecond), and the time step is set to 0.1 fs in our simulation.

To mimic the photoexcitation, we add A-field in the *k*-space of Hamiltonian.

$$H(t) = 1/2(-i\nabla + \mathbf{A}(t))^2 \tag{4}$$

For rt-TDDFT Hamiltonian, we further obtain,

$$H(t) = 1/2(-i\nabla_x + A_x \times E(t))^2 + 1/2(-i\nabla_y + A_y \times E(t))^2 + 1/2(-i\nabla_z + A_z \times E(t))^2 \tag{5}$$

In our work, we design $A_x = A_y = A_z$, so the electric field polarization is average along the *x*, *y*, and *z* directions. $E(t)$ is an external electric field of the time dimension to simulate a laser pulse with a Gaussian shape in our rt-TDDFT,

$$E(t) = E_0 \cos(\omega t) \exp[-(t-t_0)^2/(2\sigma^2)] \tag{6}$$

$E_0$ is a constant in V/Å unit. We choose two group parameters in our simulations: $t_0$ = 75 fs, $2\sigma$ = 25 fs is the pulse width, and $\omega$ = 3.2 eV is the photon energy which is consistent with a 387-nm optical pump pulse in experiment [13]; $t_0 = 50$ fs, $2\sigma = 25$ fs is the pulse width, and $\omega = 2.03$ eV is the photon energy which is consistent with a 610-nm optical pump pulse in another experiment [12].

## Note 2: Simulations of valence electrons constrained to the conduction band.

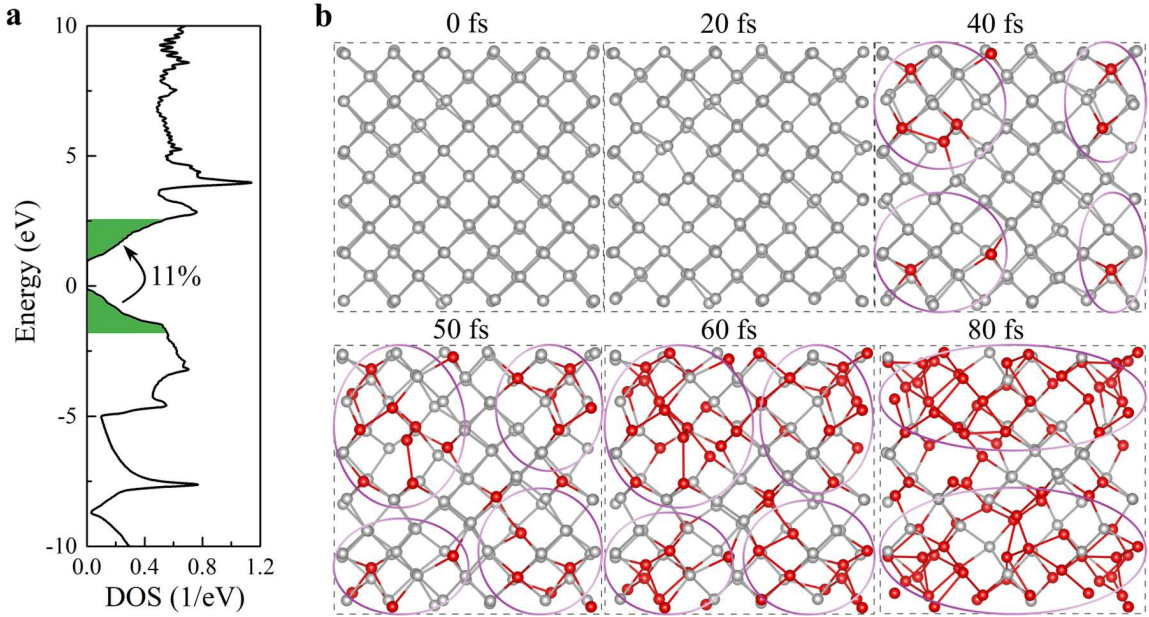

**Supplementary Figure 1 | Ultrafast dynamics with 11% of valence electrons fixed to the conduction band edge (CBE). a,** Density of states (DOS) of Si. Shaded areas represent excited hole and electron occupations. The VBM is set to 0 eV and CBM is 1.12 eV, respectively. **b,** The snapshots of atomic displacements in the $(x, y)$ plane at 0 fs, 20 fs, 40 fs, 50 fs, 60 fs, and 80 fs, respectively. The red atoms belong to the Lindemann particle $(R_i(t) - R_i(0) > 0.35$ Å$)$, which represents the molten atoms.

To further certify the homogenous nucleation mechanism, we design an initial electronic excitation state where the valence electrons near valance band edge (VBE) are moved to near conduction band edge (CBE), as shown in Supplementary Fig. 1a. Due to the PBE pseudopotential with 4 valence electrons in Si used in this study, the simulated 64-atoms system contains 256 valence electrons. Taking 11% excitations as an example, almost 28 electrons from the VBE are moved to the CBE. Based on the occupations of electronic excitations near CBE, we have re-run a simulation in a 300 K initial temperature. We find the Si crystal achieves the nonthermal melting within 80 fs. The nonthermal melting starts to occur in some local regions, then diffuses to the whole crystal, as shown in Supplementary Fig. 1b, which further proves the mechanism of local melting.

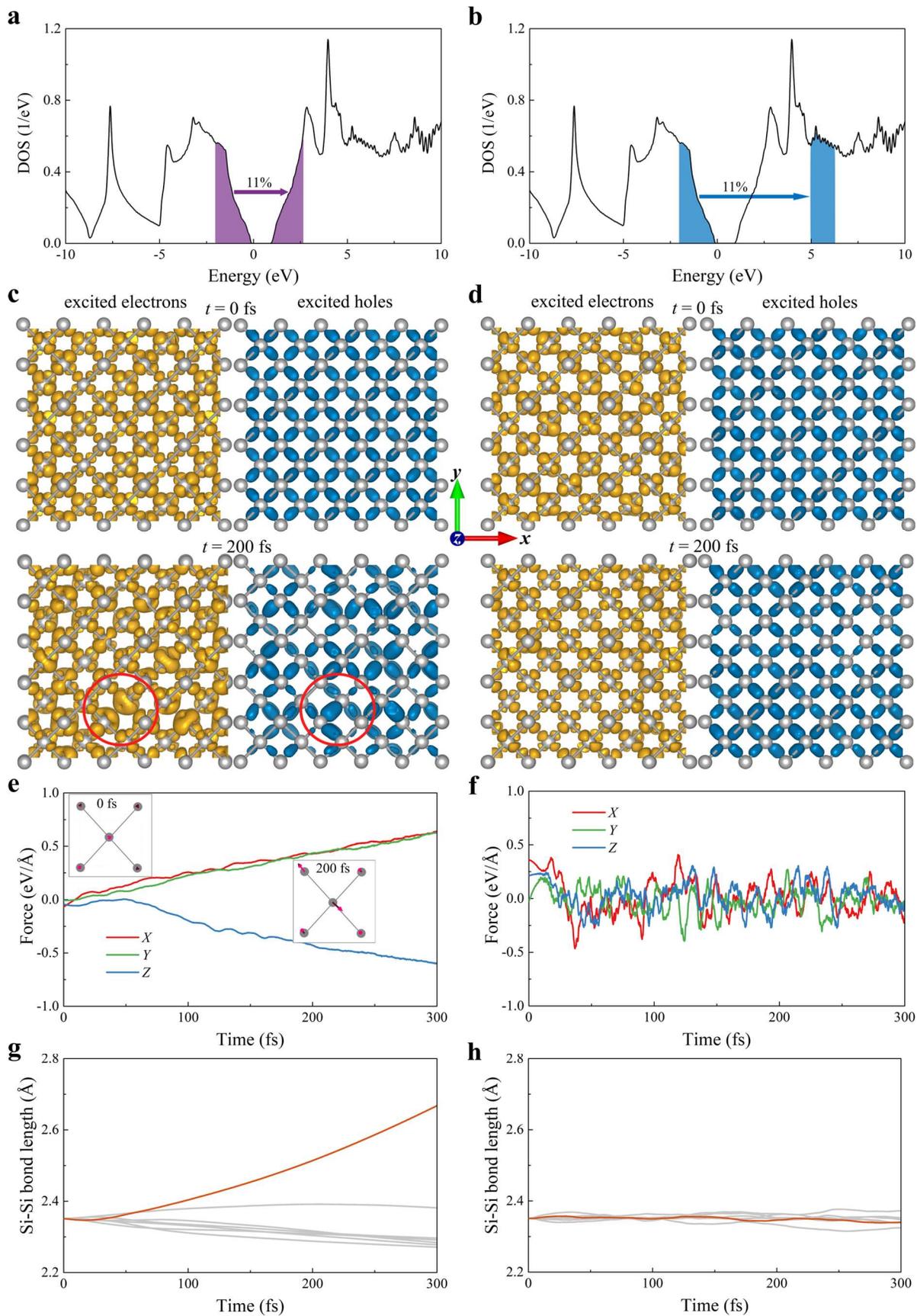

**Supplementary Figure 2 | Distribution of excited electrons (holes) and atomic driving force at an extremely low temperature (~ 1K). a, c, e, g,** for the case of valence electrons excited to lower levels; **b, d, f, h,** for the case of valence electrons excited to higher levels. **a, b,** Density of states (DOS) of Si. Shaded areas represent excited hole and electron occupations. **c, d,** Real-space distributions of excited electrons and holes at 0 fs and 200 fs. The accumulated electrons and holes at 200 fs are marked in red circles (Supplementary Fig. 2c), which induce the locally atomic distortion. **e, f,** Evolution of atomic driving force along $x$, $y$, $z$ directions. The insets (Supplementary Fig. 2e) show the driving resultant force projected in the $(x, y)$ plane at 0 fs and 200 fs, and the pink arrows represent the direction and magnitude of forces. **g, h,** Evolution of Si-Si bond length.

We constrain the valence electrons to near CBE and higher-energy levels as an initial state of electronic excitations, as shown in Supplementary Fig. 2a and 2b. To further verify the essence of nonthermal melting, we re-run the rt-TDDFT simulations at an extremely low temperature (~ 1K) to expulse the effect of thermally lattice vibrations. Initially, whether the excited electrons occupy the lower levels or the higher levels, the excited charge density is both high symmetry at t = 0 fs (Supplementary Fig. 2c and 2d), and the forces on atoms are zero (Inset in Supplementary Fig. 2e). In the case of excited electrons occupying CBE, the excited electrons and holes localize certain Si-Si bonds by charge transfer at 200 fs, marked in red circles (Supplementary Fig. 2c). The localized charge distributions trigger the microscopic driving forces along with the Si-Si bonding directions (Supplementary Fig. 2e), which induce the breaking of the Si-Si bond (Supplementary Fig. 2g). However, in the case of excited electrons occupying high-energy states, the excited carriers do not transfer certain Si-Si bonds. There are no corresponding atomic driving forces along Si-Si bonding directions (Supplementary Fig. 2f) to break the Si-Si bonds (Supplementary Fig. 2h).

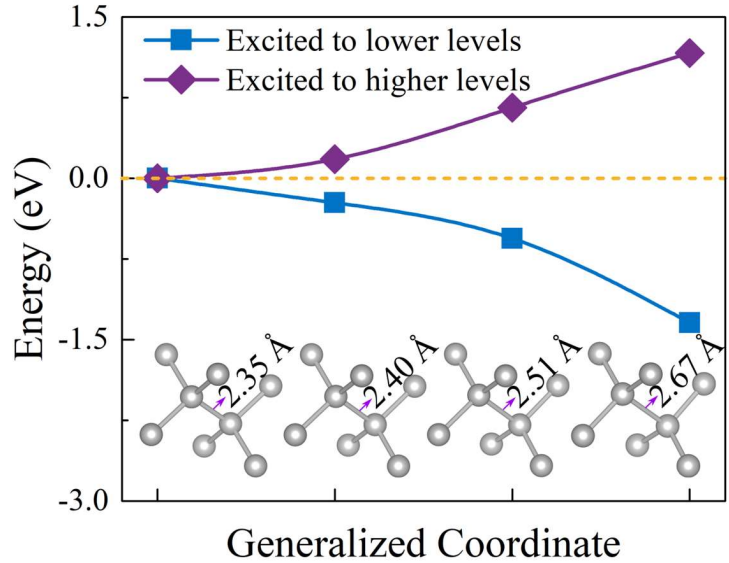

**Supplementary Figure 3 | Potential energy surfaces (PESs).** The total energy of the Si system in different excitation states as a function of Si-Si bond length.

For the 610-nm case, they are excited to these states closer to the band edge, hence close to the bonding and antibonding states, while for the 387-nm case, the initial excited electrons and holes are in much higher energy states which cannot help with the Si-Si bond weakening. To further prove the conclusion, we apply the constrained DFT to calculate the potential energy surfaces (PESs) along Si-Si bond stretching at excitation-state occupations of 610-nm and 387-nm laser pumping, respectively (Supplementary Fig. 3). We find that occupying the electron states according to the 610-nm case can indeed reduce the local PES. But occupying the electron states according to the 387-nm case, the PES is hardly any effect. This means that ultrafast nonthermal melting is not manipulated by the increasing energy pumped into the system, rather the occupations of the special states (band-edge states) that are important for nonthermal melting.

According to the occupations of excited electrons from 610-nm and 387-nm laser pumping, we utilize constrained DFT to calculate the potential energy surfaces (PESs) along with Si-Si bond stretching (Supplementary Fig. 3). In the case of electrons occupying lower-energy levels (610-nm laser), the total energy gradually reduces along the Si-Si stretching direction which illustrates the initial ideal structure is unstable (Supplementary Fig. 3). In the case of electrons occupying higher energy levels, the total energy remains a minimum value at the ideal structure (Supplementary Fig. 3). By comparing the change PESs at different electronic occupations, we verify our conclusion again.

## Note 3: Boltzmann-TDDFT methods

As discussed above, the original rt-TDDFT evolution does not satisfy the detailed balance, hence cannot be used to describe hot carrier cooling.[57-59] The detailed balance is based on adiabatic states. Since we expand our wave function with the adiabatic states as shown in Eq.(2), this provides a unique opportunity for us in dealing with the detailed balance. To introduce detailed balance, we first need to define a charge flow between adiabatic states $i$ and $i'$. This can be defined as:

$$T(i, i', t) = -\sum_{j=1}^{N} 2Re\{iC_{j,i}^*(t)V_{i,i'}(t)C_{j,i'}(t)\} \quad (8)$$

Here $V$ is the quantity in Eq.(3), and $C_{j,i}$ is the wave function expansion coefficient. This $T(i, i', t)$ describes the charge flow from adiabatic state $i'$ to adiabatic state $i$. This charge flow comes from all wave functions $\psi_j(t)$, not just from any one of them. The $T(i, i', t)$ is invariant under unitary rotations in the occupied subspace of $\psi_j(t)$. Note: $T(i, i', t) = -T(i, i', t)$. To introduce the decoherence effect, we first define time averaged $T(i, i', t)$ as:

$$I(i, i', t) = \frac{1}{\tau_{i,i'}} \int_0^\infty T(i, i', t - t') e^{-\frac{t'}{\tau_{i,i'}}} dt' \quad (9)$$

Here $\tau_{i,i'}$ is the decoherence time between adiabatic states $i$ and $i'$. Now, to restore the detailed balance, we like to change the time evolution Eq.(3), so the averaged charge flow from $i'$ to $i$ will be altered according to the detailed balance. To satisfy the detailed balance, we need to modify $I(i, i', t)$ by adding a $\Delta I(i, i', t)$ as:

$$\Delta I(i, i', t) = \begin{cases} I(i, i', t)(e^{-|\epsilon_i - \epsilon_{i'}|/kT} - 1), & I(i, i', t)(\epsilon_i - \epsilon_{i'}) > 0 \\ 0, & I(i, i', t)(\epsilon_i - \epsilon_{i'}) \leq 0 \end{cases} \quad (10)$$

Here $\epsilon_i$, $\epsilon_{i'}$ are the adiabatic eigenstates, $T$ is the temperature. Thus, $\Delta I(i, i', t)$ can be considered as the correction to the charge flow $T(i, i', t)$. This correction can be realized by modifying the wave function $C_{j,i}(t)$ by adding a $\Delta C_{j,i}(t)$ after every wave function evolution step from $t_1$ to $t_1+\Delta t$.

$$\sum_{i=1}^{M} C_{j_1,i}(t) \Delta C_{j_2,i}^*(t) + \Delta C_{j_1,i}(t) C_{j_2,i}^*(t) = 0 \quad (11)$$

$$2\sum_{j=1}^{N} Re[C_{j,i}(t)\Delta C_{j,i}^*(t)] = \Delta t \sum_{i'} \Delta I(i, i', t) \quad (12)$$

The first equation is used to satisfy the orthonormal condition for $\psi_j(t)$, while the second equation is used to modify the occupation of the adiabatic states by introducing $\Delta I(i, i', t)$. The above linear equation is solved by the conjugate gradient method. Note that, usually there are more unknown parameters than the number of equations. Thus, minimum amplitude $\Delta C_{j,i}(t)$ solution is sought which satisfies the above equations. After the introduction of $\Delta C_{j,i}(t)$, the energy is not conserved. The energy conservation is restored by subtracting a velocity in the transition degree of freedom, which can be calculated from $\Delta I(i, i', t)$.[46]

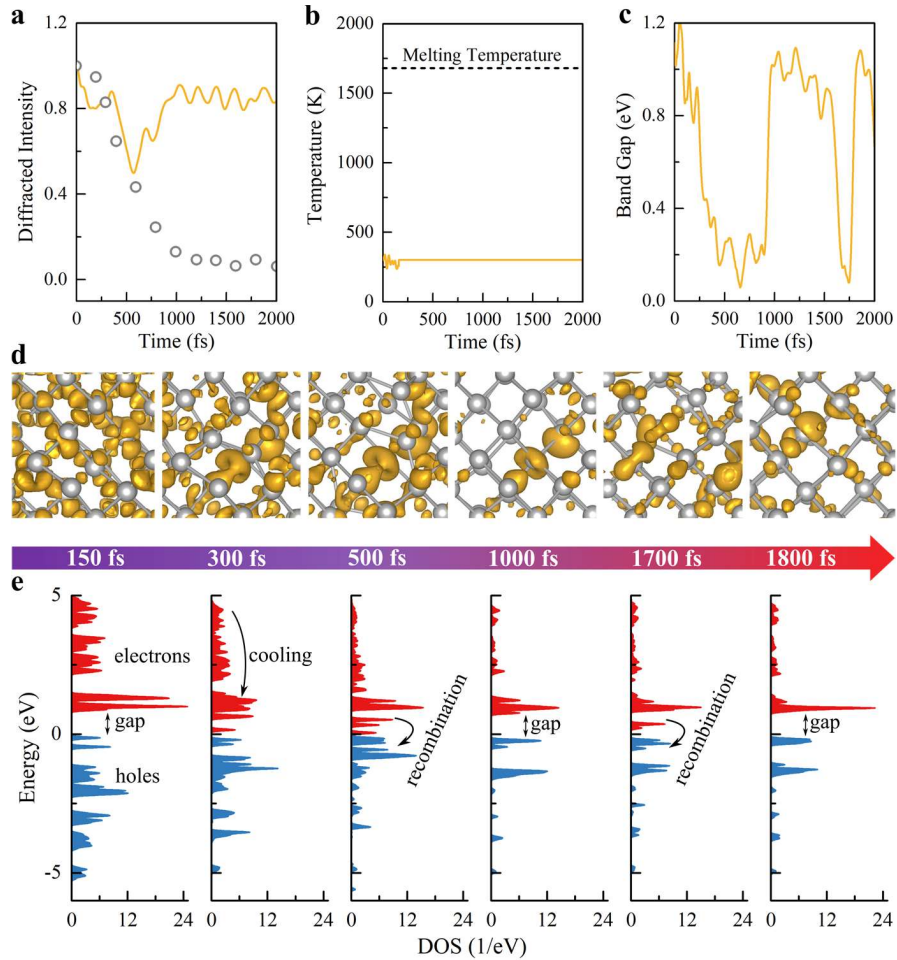

**Supplementary Figure 4 | Role of hot carrier cooling on Si structural nonthermal melting. a,** The simulated electron diffracted intensity of Si (220) Bragg peaks after 387-nm laser pulse pumping. The experimental data (gray circles) are from Si (220) peak with 11% electrons excited by 387-nm 150-fs pump

pulses. [13] **b,** Evolution of crystal temperature versus time. **c,** Evolution of band gap versus time. **d,** Real-space distributions of excited electrons (yellow color) and atomic structures (silver balls) at different times. **e,** for distributions of excited carriers as a function of time. Red (blue) shaded areas represent excited electron (hole) occupations.

Based on the new Boltzmann-TDDFT method, we have well described the effect of hot carrier cooling. In the NVE ensemble, the total energy is a constant, and the hot carriers will transfer extra energy to cold lattice through electron-phonon (*el-ph*) coupling to increase the system temperature (Fig. 4). Besides, we can remove this hot carrier cooling to cause the increase of the lattice temperature lattice by performing an NVT simulation, where the kinetic energy of the lattice is rescaled to be kept at 300 K (Fig. 4b). We illustrate that the carrier distributions induced by hot carrier cooling play a crucial role in structural nonthermal melting. As we see in Supplementary Fig. 4a, initially, the system is still broken around 500 fs (locally nonthermal distortion in Supplementary Fig. 4) by carrier cooling to band edge (Supplementary Fig. 4e). However, after that, as the electrons and holes fast recombine due to the disappearance of the bandgap (Supplementary Fig. 4c), the excited electrons go back to the ground state to reduce the driving force on atoms which results in the system returning to crystal structure at 1000 fs in Supplementary Fig. 4d.

**Reference**

1  Harb, M. *et al.* Electronically driven structure changes of Si captured by femtosecond electron diffraction. *Phys Rev Lett* **100**, 155504, doi:10.1103/PhysRevLett.100.155504 (2008).
2  Tom, H. W., Aumiller, G. D. & Brito-Cruz, C. H. Time-resolved study of laser-induced disorder of Si surfaces. *Phys Rev Lett* **60**, 1438-1441, doi:10.1103/PhysRevLett.60.1438 (1988).
3  Ren, J., Vukmirović, N. & Wang, L.-W. Nonadiabatic molecular dynamics simulation for carrier transport in a pentathiophene butyric acid monolayer. *Physical Review B* **87**, doi:10.1103/PhysRevB.87.205117 (2013).
4  Parandekar, P. V. & Tully, J. C. Detailed Balance in Ehrenfest Mixed Quantum-Classical Dynamics. *J Chem Theory Comput* **2**, 229-235, doi:10.1021/ct050213k (2006).
5  Parandekar, P. V. & Tully, J. C. Mixed quantum-classical equilibrium. *J Chem Phys* **122**, 094102, doi:10.1063/1.1856460 (2005).
6  Wang, L. W. Natural Orbital Branching Scheme for Time-Dependent Density Functional Theory Nonadiabatic Simulations. *J Phys Chem A* **124**, 9075-9087, doi:10.1021/acs.jpca.0c06367 (2020).